\documentstyle[psfig]{article}
\begin{document}
\LARGE
\begin{center}
Lines on Del Pezzo surfaces and transfinite heterotic string spacetimes
\end{center}
\vspace*{.2cm}
\Large
\begin{center}
Metod Saniga

\vspace*{.3cm}
\small
{\it Astronomical Institute, Slovak Academy of Sciences, 
SK-059 60 Tatransk\' a Lomnica, Slovak Republic}
\end{center}

\vspace*{-.4cm}
\noindent
\hrulefill

\vspace*{.2cm}
\small
\noindent
{\bf Abstract}\\

It  is  pointed  out  that the hierarchy of fractal dimensions characterizing
transfinite  heterotic  string spacetimes bears a striking resemblance to the
sequence of the number of lines lying on Del Pezzo surfaces.

\noindent
\hrulefill

\vspace*{.5cm}
\normalsize
Employing the notion and properties of the so-called Cantorian fractal space,
${\cal E}^{(\infty)}$, El Naschie has recently  demonstrated  [1--6] that the  
transfinite  heterotic string spacetimes are  endowed with the following five
characteristic {\it fractal} dimensions
\begin{equation}
D_{\kappa}=\frac{(\alpha_{0}^{-1}) \phi^{2+\kappa}}{\langle d_{c}^{(2)} 
\rangle}=\frac{\bar{\alpha}_{0}}{2} \phi^{2+\kappa},
\end{equation}
where  $\kappa=0,1,\ldots,4$, $\bar{\alpha}_{0}$  is the inverse value of the 
fine structure constant and  $\phi$ stands for the Hausdorff dimension of the 
kernel Cantor set. Taking $\bar{\alpha}_{0} \simeq 137.082$ and $\phi= 1 -
\phi^{2} = \left(
\sqrt{5} - 1 \right)/2 \simeq 0.618034$,  he  found  the following remarkable 
series [1,6]
\begin{equation}
D_{0} \simeq 26.18034 ~\rightarrow~ D_{1} \simeq 16.18034 ~\rightarrow~ D_{2} 
\simeq 10 ~\rightarrow~ D_{3} \simeq 6.18034 ~\rightarrow~ D_{4} \simeq 
3.81966.
\end{equation}
The aim of this short contribution is to show that this sequence, in its {\it 
integer}-valued part, can almost exactly be reproduced by the arrangement  of  
the multiplicities of the configurations of lines lying on Del Pezzo surfaces.

A Del Pezzo surface [7],  $F_{n}$, is a rational surface of order $n$ sitting 
in an $n$-dimensional projective space, where $n$=3,4,\ldots,9. Taken together, 
these   surfaces  form  a  single  simple  series  $F_{9}  \rightarrow  F_{8} 
\rightarrow \ldots \rightarrow F_{3}$,  such that $F_{n}$ ($3 \leq n \leq 8$) 
is  always  the  projection of $F_{n+1}$ from a point of itself. In addition, 
all of them can be  represented  on a projective plane by means of systems of 
non-singular cubic curves  having $9-n$ base (i.e. shared by all the members)
points. A line of $F_{n}$ is  mapped on the plane  into  a base point, a line 
joining two base points, or a  conic passing through five  of the base points 
(see, e.g. [8]);  hence,  the  number of lines lying on $F_{n}$, $\Theta(n)$, 
is simply 
\begin{equation}
\Theta(n)=(9-n)+\left(\hspace*{-.2cm}\begin{array}{c} 9-n \\ 2 \end{array}
\hspace*{-.2cm}\right)+ \left(\hspace*{-.2cm}\begin{array}{c} 9-n \\ 5
\end{array}\hspace*{-.2cm}\right),
\end{equation}
with the understanding that $\left(\hspace*{-.1cm}\begin{array}{c} a \\ b 
\end{array}\hspace*{-.1cm}\right) 
\equiv 0$ if $a < b$. In particular, for $3\leq n \leq7$ we have
\begin{equation}
\Theta(3)=27 ~\rightarrow~ \Theta(4)=16 ~\rightarrow~ \Theta(5)=10 
~\rightarrow~ \Theta(6)=6 ~\rightarrow~ \Theta(7)=3,
\end{equation}
which, except for the first term, is indeed seen to be an exact integer-valued 
match  for  the  fractal  sequence given by Eq. (2). It is also worth noticing 
that
\begin{equation}
\Theta(3)-\Theta(4)=27-16=11=10+1=\Theta(5)+1
\end{equation}
and
\begin{equation}
\Theta(5)-\Theta(6)=10-6=4=3+1=\Theta(7)+1.
\end{equation}

In order to better understand the origin of this hierarchy, as well as to see 
how intricate the connection between the individual Del Pezzo surfaces is, we 
project  $F_{n}$,  $n \geq 4$, into a three-dimensional projective  space [7], 
denoting  these  projected  surfaces  as $\widehat{F}_{n}$. We first take our 
familar  cubic  surface  $F_{3} \equiv \widehat{F}_{3}$, and the twenty-seven 
lines on it [9].  If  we  disregard  any one line and the ten lines which are 
incident  with it,  then  the {\it sixteen} remaining lines are, as for their 
mutual  intersections,  related  to  each other as the sixteen lines lying on 
$\widehat{F}_{4}$.  Analogously,  if  on  $\widehat{F}_{4}$ we ignore any one 
line  and the five lines that meet it, the {\it ten} remaining lines have the 
same intersection properties as the ten lines on $\widehat{F}_{5}$. Similarly, 
if on $\widehat{F}_{5}$ we omit one line and the three lines incident with it, 
we are left with {\it six} lines exhibiting the same algebra as the six lines 
situated  on $\widehat{F}_{6}$. And finally, if on $\widehat{F}_{6}$ we leave 
out  any  one  line  and the two lines that meet it, the configuration of the 
{\it  three}  remaining lines enjoys the same properties as that of the three 
lines upon $\widehat{F}_{7}$. At this point the procedure becomes unambiguous 
as an  $\widehat{F}_{7}$ contains {\it two} different sets of skew lines; one 
featuring one line and the other being empty. There are, therefore, two kinds
of the 3-D projected and, hence, also parent Del Pezzo surfaces of order eight 
[8]:  a  single-line octavic surface, $F_{8}$, belonging to the main sequence 
discussed above  and a line-free octavic surface, $F_{8}^{*}$, which lies off 
the main sequence  since it {\it cannot} be represented on the plane in terms 
of cubic curves. Expressed schematically
\begin{eqnarray}
F_{9} \rightarrow F_{8} \hspace*{4.8cm} \nonumber \\
\searrow \hspace*{4.2cm} \nonumber \\
F_{7} \rightarrow F_{6} \rightarrow F_{5} \rightarrow F_{4} \rightarrow 
F_{3}. \\
\nearrow \hspace*{4.2cm} \nonumber \\
F_{8}^{*} \hspace*{4.8cm} \nonumber
\end{eqnarray}
And  it  is  this `branching' of the Del Pezzo sequence at $n=7$ that may, in 
our  opinion,  account  for the fact why the above described amazing parallel  
with the dimensionalities of heterotic string spacetimes ends at $\kappa=4$.

It is of crucial importance to  observe next that the above described analogy
admits  an  intriguing  extension  at  the  opposite  end of the sequence. By 
projecting $F_{3}$ from its generic point onto a plane we obtain the so-called 
Del  Pezzo  double-plane $\widetilde{F}_{2}$ [8]. This rational surface is an  
improper, yet natural member of the series. It can be viewed  as  a  pair  of  
superimposed  planes  which `touch' each other along a branch curve $\Gamma$;  
the latter is a non-singular quartic curve, being the projection of the curve 
of contact of the proper tangent cone to $F_{3}$ from the point of projection.
It  is  obvious  that  $\widetilde{F}_{2}$  is  birationally representable on 
a(n ordinary) plane by means of an aggregate of non-singular cubics featuring
$9-2=7$ base points. From Eq. (3) we then find that the surface  $\widetilde{
F}_{2}$ possesses 7 lines that correspond to the base points, 21 lines  which  
have  their  counterparts in the lines joining the base points in pairs,  and  
another 21 lines answering to the conics  passing  through quintuplets of the 
base points. It, however, contains additional 7 lines, the ones having  their
images  in nodal cubic curves passing once through six of the base points and  
twice via the remaining seventh base point.  Hence,  a Del Pezzo double-plane 
contains 56 lines altogether, and these form 28 dual superimposed pairs lying 
along   the   28  bitangents  of  $\Gamma$   [8].  And  there  does  exist  a 
56=28+28-dimensional algebra of superstrings [10]. It is nothing but an $N=8$ 
theory in the ordinary four-dimensional spacetime;  the theory is obtained by 
compactifying 11-dimensional supergravity down to  four dimensions on $T_{7}$ 
and is $S$-dual to type IIA string theory compactified on $T_{6}$. It features 
exactly  56  central  charges, which are arranged in 28 dual pairs; for there 
are  28  vector  gauge  fields  in  total, and each of them is coupled to one  
electric  and one magnetic charge [10].  Out of them, 7 pairs result from the 
supersymmetry algebra itself, while the remaining 21 doublets are obtained by 
wrapping a membrane and  a  5-brane,  respectively;   and  this factorization 
corresponds  exactly  to  the  above-described  $7 + 21 + 21 + 7$ grouping of 
the  lines  on $\widetilde{F}_{2}$,  resulting  from  the  particular  method  
of their representation! As for the transfinite case, we incidentally note 
that the dimension in question can be retrieved as follows [11]
\begin{equation}
D_{0} + D_{1} + D_{2} + D_{4} = D_{0} \left(2+\phi^{4}\right)
\simeq 56.18034 \simeq 10 \otimes (5 + \phi).
\end{equation}

All the above-introduced facts and findings thus cast completely new light on 
our  hypothesis  put  forward  in  [9],  which  now  allows  of the following 
intriguing  extension  and  generalization:  the  hierarchy of characteristic 
dimensions  exhibited  by  the  transfinite heterotic string spacetimes seems
to  be  strongly  underpinned  by  the algebra of the configurations of lines 
lying on Del Pezzo surfaces of order two to seven. 
\\ \\ \\
{\bf Acknowledgements}
\\ \\
I am indebted to Prof. El Naschie (Cambridge University) for drawing my
attention to the 56-dimensional representation of superstrings discussed
above. This work was supported in part by the NATO Collaborative Linkage
Grant PST.CLG.976850.
\\ \\ \\ \\

\end{document}